\documentclass[copyright,creativecommons]{eptcs}
\providecommand{\event}{FICS 2013} 
\usepackage{breakurl}             

\usepackage{etoolbox}
\newbool{full}
\boolfalse{full}

\newcommand{\takeout}[1]{\empty}

\usepackage[final,inline]{fixme}

\usepackage{enumerate} 

\usepackage{amsmath,amsthm,eufrak}
\usepackage{hyperref}
\usepackage{array}
\usepackage{multirow}
\usepackage{color}
\usepackage{mathrsfs}
\usepackage{relsize}
\usepackage{lscape}
\usepackage{
pgfrcs}
\usepackage{
pgf}
\usepackage{
tikz}
\usetikzlibrary{%
  shapes.misc,
  shapes.geometric,
  positioning,
  shadows%
}
\usepackage[all]{xy}
\xyoption{2cell}
\xyoption{curve}
\UseTwocells
\SelectTips{cm}{}
\usepackage{stmaryrd}
\usepackage{pigpen}  

\newtheorem{theorem}{Theorem}[section]
\newtheorem{proposition}[theorem]{Proposition}
\newtheorem{corollary}[theorem]{Corollary}
\newtheorem{lemma}[theorem]{Lemma}
\theoremstyle{definition}
\newtheorem{definition}[theorem]{Definition}
\newtheorem{assumption}[theorem]{Assumption}
\newtheorem{example}[theorem]{Example}
\newtheorem{remark}[theorem]{Remark}
\newtheorem{examples}[theorem]{Examples}
\newtheorem{construction}[theorem]{Construction}

\numberwithin{equation}{section}

\newcommand{\modcol}[1]{{\color{blue}#1}}

\newcommand{\ibox}{\modcol{\rightslice}}

\newcommand{\unpdg}{\ddagger} 

\newcommand{\curry}[3]{\mathsf{curry}^{#1}_{#2,#3}}
\newcommand{\uncurry}[3]{\mathsf{uncurry}^{#1}_{#2,#3}}
\newcommand{\eval}[2]{\mathsf{eval}_{#1,#2}}
\newcommand{\iprod}[2]{\mathsf{can}^{-1}_{#1,#2}}

\newcommand{\WG}{\mathfrak{W}}

\newcommand{\presh}[2]{\mathsf{presh}(#1,#2)}

\newcommand{\bxtrm}{\modcol{\mathfrak{b}}\!}
\newcommand{\fxtrm}{\modcol{\mathfrak{f}}}
\newcommand{\dltrm}{\modcol{\mathfrak{p}}\!}

\newcommand{\dfsp}{\;}

\newcommand{\comono}[3]{\bxtrm\dfsp#1 [\ibox#2\modcol{\leadsto}#3]}

\newcommand{\fix}[2]{\fxtrm#1.\dfsp#2} 
\newcommand{\delay}[1]{\dltrm\dfsp#1}

\newcommand{\deq}{:=} 

\newcommand{\cat}[1]{\mathcal{#1}}

\newcommand{\catC}{\cat{C}}
\newcommand{\catD}{\cat{D}}

\def\A{\cat A}
\def\C{\catC}
\def\D{\catD}
\def\cpo{\mathsf{CPO}}
\def\cpob{\mathsf{CPO}_\bot}
\def\cms{\mathsf{CMS}}

\def\setc{\mathsf{Set}}

\def\refeq#1{(\ref{#1})}
\def\prl{\pi_\ell}
\def\prr{\pi_r}
\def\can{\mathsf{can}}
\def\Id{\mathsf{Id}}
\def\ol#1{\overline{#1}}
\def\sol#1{#1^\dagger}
\def\ssol#1{#1^\unpdg}
\def\Tr{\mathsf{Tr}}
\def\op{\mathsf{op}}

\def\subto{\hookrightarrow}
\def\inr{\mathsf{inr}}
\def\inl{\mathsf{inl}}
\def\can{\mathsf{can}}

\newcommand{\tragger}{{\Tr_\dagger}}
\newcommand{\dace}{{\dagger_\Tr}}

\newcommand{\tlnt}[1]{\tlnote[inline,marginclue]{#1}}
\newcommand{\smnt}[1]{\smnote[inline,marginclue]{#1}}

\FXRegisterAuthor{sm}{asm}{SM}
\FXRegisterAuthor{tl}{atl}{TML}
\title{Guard Your Daggers and Traces: On The Equational Properties of Guarded (Co-)recursion}

\author{Stefan Milius \quad\qquad\qquad Tadeusz Litak
\institute{Chair for Theoretical Computer Science (Informatik 8)}
\institute{Friedrich-Alexander University Erlangen-N\"{u}rnberg, Germany}
\email{\quad mail@stefan-milius.eu \ \qquad tadeusz.litak@gmail.com}
}
\def\titlerunning{On The Equational Properties of Guarded (Co-)recursion}
\def\authorrunning{S.~Milius and T.~Litak} 

\begin{document}

\maketitle

\begin{abstract}
Motivated by the recent interest in models of guarded (co-)recursion we
study its equational properties. We formulate axioms for guarded fixpoint operators
generalizing the axioms of iteration theories of Bloom and \'Esik.  Models of these axioms include both standard (e.g., cpo-based) models of iteration theories and models of guarded recursion such as complete metric spaces or the topos of trees
studied by Birkedal et al. We show that the standard result  on the satisfaction of all  Conway axioms by a unique dagger operation generalizes to the guarded setting. We also introduce the notion of guarded
trace operator on a category, and we prove that guarded trace and
guarded fixpoint operators are in one-to-one correspondence. Our
results are intended as first steps leading to the description of classifying
theories for guarded recursion and hence completeness results
involving our axioms of guarded fixpoint operators in future work. 
 \end{abstract}

\section{Introduction}

Our ability to  describe concisely potentially infinite
computations or infinite behaviour of systems relies on 
recursion, corecursion and iteration. Most programming
languages and specification formalisms include a fixpoint
operator. In order
to give semantics to such operators one usually considers either
\begin{itemize}
\item models based on complete partial orders where fixpoint operators are
interpreted by  least fixpoints using the Kleene-Knaster-Tarski
theorem or
\item models based on complete metric spaces and unique
fixpoints via Banach's theorem or 
\item  term models where unique
fixpoints arise by unfolding specifications
syntactically. 
\end{itemize}

In the last of these cases, 
one only considers \emph{guarded} (co-)recursive definitions;  see
e.g.~Milner's solution theorem for CCS~\cite{milner89} or Elgot's
iterative theories~\cite{elgot75}. 
Thus, the fixpoint operator becomes a
partial operator defined only on a special class of maps. For a
concrete example consider complete metric spaces which form a category
with all non-expansive maps as morphisms, but unique fixpoints are
taken only of contractive maps.

Recently, there has been a wave of interest in expressing guardedness by a
new type constructor $\ibox$, a kind of ``later'' modality, which
allows to make the fixpoint operator total, see, e.g.,
 Nakano~\cite{Nakano00:lics,Nakano01:tacs}, Appel et al. \cite{AppelMRV07:popl}, Benton and
Tabareau~\cite{BentonT09:tldi}, Krishnaswami and
Benton~\cite{KrishnaswamiB11:lics,KrishnaswamiB11:icfp}, Birkedal et
al.~\cite{BirkedalMSS12:lmcs,BirkedalM13:lics} and Atkey and
McBride~\cite{AtkeyMB13:icfp}.  For example, in the case
of complete metric spaces $\ibox$ can be  an endofunctor scaling the metric
of any given space by a fixed factor $0<r<1$ so that non-expansive
maps of type $\ibox X \to X$ are precisely contractive maps with a
contraction factor of at most $r$. This allows to define a guarded (parametrized) fixpoint
operator on \emph{all} morphisms of type $\ibox X \times Y
\to X$ of the model. So far various models allowing the interpretation
of a typed language including a guarded fixpoint operator have been
studied:  complete metric spaces,  the ``topos of
trees'', i.e.,\ presheaves on $\omega^{op}$~\cite{BirkedalMSS12:lmcs}
or, more generally, sheaves on complete Heyting
algebras with a well-founded basis~\cite{DiGianantonioM04:fossacs,BirkedalMSS12:lmcs}.

This paper initiates the study of the essential 
properties of guarded fixpoint operators. In the realm of ordinary
fixpoint operators, it is well-known that  iteration
theories of Bloom and \'Esik~\cite{be93} completely axiomatize
equalities of fixpoint terms in models based on complete partial
orders (see also Simpson and Plotkin~\cite{sp00}). We make here the
first steps towards similar completeness results in the guarded setting. 

We begin with formalizing the notion of guarded fixpoint operator on a cartesian category. We discuss a number of models, including not only all those mentioned above, but also some  not mentioned so far in the context of $\ibox$-guarded (co-)recursion. 
In fact, we consider the inclusion of examples such as the lifting functor on $\cpo$ (which also happens to be a paradigm example of a \emph{fixpoint monad}, see Example \ref{ex:cats}.\ref{ex:cpo} and the concluding remark of Section \ref{rem:monads}) or completely iterative monads (see Section \ref{sec:cim}) a pleasant by-product of our work and a potentially fruitful connection for future research. Then, we formulate generalizations of standard
iteration theory axioms for guarded fixpoint operators and we
establish these axioms are sound in all models under consideration. In particular, the central result of Section~\ref{sec:fix} is 
Theorem~\ref{thm:unique}: models with \emph{unique} guarded fixpoint
operators satisfy all our axioms. 


Hasegawa~\cite{h97} proved that giving a parametrized fixpoint operator on a category satisfying
the so-called \emph{Conway axioms} (see, e.g.,~\cite{be93,sp00} and Section \ref{sec:prop}
 below) is
equivalent to giving a \emph{traced cartesian structure} \cite{jsv96} on that
category.\footnote{\emph{Cartesian} here refers to the monoidal
  product being the ordinary categorical product.} Section~\ref{sec:tr} lifts this result to the guarded setting. We introduce a natural notion of a guarded trace operator
on a category, and we prove in Theorem~\ref{thm:tr} that guarded traces
and guarded fixpoint operators are in one-to-one correspondence.
This extends to an isomorphism between the (2-)categories of guarded traced
cartesian categories and guarded Conway categories.

Section~\ref{sec:conc} concludes and discusses  further work. 

Proofs of the major theorems 
will be made available in the full version.

\subsection{Notational conventions}

We will assume throughout that readers are familiar with basic notions
from category theory. We denote the product of two objects by
\[
\xymatrix@1{
  A & A \times B \ar[l]_-{\prl} \ar[r]^-{\prr} & B,
  }
\]
and $\Delta: A \to A \times A$ denotes the diagonal. For every functor
$F$ we write $\can = \langle F\prl, F\prr\rangle: F(A\times B) \to FA
\times FB$ for the canonical morphism. 

We denote by $\cpo$ the
category of complete partial orders (cpo's), i.e.\ partially ordered sets (not
necessarily with a least element) having joins of $\omega$-chains. The
morphisms of $\cpo$ are Scott-continuous maps, i.e.\ maps preserving
joins of $\omega$-chains. And $\cpob$ is the full subcategory of
$\cpo$ given by all cpo's with a least element $\bot$. We will also
consider the category $\cms$ of complete 1-bounded metric spaces and
non-expansive maps.

\section{Guarded Fixpoint Operators}
\label{sec:fix}

In this section we define the notion of a guarded fixpoint
operator on a cartesian category and  present an extensive list of examples. Some of these examples like  the lifting functor $(-)_\bot$ on $\cpo$ (see Example \ref{ex:cats}.\ref{ex:cpo}) or completely iterative monads (see Section \ref{sec:cim}) do not seem to have been considered as instances of the guarded setting before. We
then introduce (equational) properties of guarded fixpoint
operators. These properties are motivated by and closely resemble properties of the
fixpoint operator in iteration theories of Bloom and \'Esik~\cite{be93}. 
We conclude this section with Theorem \ref{thm:unique} stating that unique fixpoint
operators satisfy all the properties we study.

\subsection{Definition and Examples of Guarded Fixpoint Operators}

\begin{assumption} \label{mainassumption}
  We assume throughout the rest of the paper that $(\catC,\ibox)$ is a pair consisting of a category $\catC$ with finite
  products (also know as a \emph{cartesian category}) and a pointed endofunctor $\ibox: \catC \to
  \catC$, i.e.\ we have a natural transformation $p: \Id \to
  \ibox$. The endofunctor $\ibox$ is called \emph{delay}.
\end{assumption}

\begin{remark}
In references like \cite{BirkedalMSS12:lmcs,BirkedalM13:lics}, much more is assumed about both the underlying category and the delay endofunctor. Whenever one wants to model simply-typed lambda calculus, one obviously imposes the condition of being cartesian closed. Furthermore, whenever one considers dependent types, one wants to postulate conditions like being a \emph{type-theoretic fibration category} (see, e.g., \cite[Definition IV.1]{BirkedalM13:lics}).   In such a case, one also wants to impose some limit-preservation or at least finite-limit-preservation condition on the delay endofunctor, see \cite[Definition 6.1]{BirkedalMSS12:lmcs}---e.g., to ensure the transfer of the guarded fixpoint operator to slices. We do not impose any of those restrictions 
 because we do not need them in this paper. It is an interesting fact that all our derivations require no more than Assumption \ref{mainassumption}. For more on the connection with the setting of \cite{BirkedalMSS12:lmcs}, see Proposition \ref{prop:mgrt} below.
\end{remark}


\begin{definition}
  \label{def:dagger}
  A \emph{guarded fixpoint operator} on $(\catC,\ibox)$ is a family of
  operations
  \[
  \dagger_{X,Y} : \catC(\ibox X \times Y, X) \to \catC(Y,X)
  \]
  such that for every $f: \ibox X \times Y \to X$ the following square
  commutes\footnote{Notice that we use the convention 
      of simply writing objects to denote the identity morphisms on them.}:
    \begin{equation}\label{eq:fixp}
      \vcenter{
        \xymatrix@C+1pc{
          Y 
          \ar[r]^-{\sol f}
          \ar[d]_{\langle \sol f, Y\rangle }
          &
          X
          \\
          X \times Y
          \ar[r]_-{p_X \times Y}
          &
          \ibox X \times Y
          \ar[u]_{f}
        }
      }
    \end{equation}
    \iffull
    \[
    \inferrule{\Gamma, x: \ibox X \vdash F: X}{\Gamma \vdash \fix{x}{F} =  F [\delay{\fix{x}{F}}/x]}
    \]
    \fi
    where (as usual) we drop the subscripts and write $\sol f:Y \to X$ in lieu of
    $\dagger_{X,Y}(f)$. We call the triple  $(\catC,\ibox, \dagger)$ a \emph{guarded fixpoint category}.
\end{definition}

Usually, one either assumes that $\dagger$ satisfies further
properties or even that $f^\dagger$ is unique such
that~\refeq{eq:fixp} commutes. 
We will come to the study of properties of guarded
fixpoint operators in Section \ref{sec:prop}. Let us begin with a list of examples.


\begin{examples}\label{ex:cats}
  \begin{enumerate}[(1)]
    \item \label{ex:identity} Taking as $\ibox$ the identity functor on $\catC$ and $p_X$
      the identity on $X$ we arrive
      at the special case of categories with an ordinary fixpoint
      operator $\catC(X \times Y, X) \to \catC(Y,X)$ (see e.g.\
      Hasegawa~\cite{h97,h99} or Simpson and Plotkin~\cite{sp00}). Concrete
      examples are: the category $\cpob$ with its usual least fixpoint
      operator or (the dual of) any iteration theory of Bloom and
      \'Esik~\cite{be93}.  
   
    \item \label{ex:constant} Taking $\ibox$ to be the constant functor on the terminal
      object $1$ and $p_X = \mathord{!}: X \to 1$ the unique morphism, a trivial
      guarded fixpoint operator is given by the family of identity
      maps on the hom-sets $\catC(Y,X)$.  
      
    \item \label{ex:cms} Take $\catC$ to be the category $\cms$ of complete 1-bounded metric spaces  (see \cite{KrishnaswamiB11:lics,KrishnaswamiB11:icfp} or \cite[Section 5]{BirkedalMSS12:lmcs} and references therein), $\ibox_r$ ($0 < r < 1$) to be an endofunctor which keeps the carrier of the space and multiplies all distances by $r$ and $p_X: X \to \ibox_r X$ to be the obvious ``contracted identity'' mapping. Note that a non-expansive mapping $f: \ibox_rX \to X$ is the same as an \emph{$r$-contractive} endomap, i.e.\ an endomap satisfying $d(fx,fy) \leq r \cdot d(x,y)$. A guarded fixpoint operator is given by an application of Banach's unique fixpoint theorem: for every $f: \ibox_r X \times Y \to X$ we consider the map 
\[
\Phi_f: \cms(Y,X) \to \cms(Y,X), \qquad \Phi_f(m) = f \cdot (p_X \times Y) \cdot \langle m, Y\rangle;
\]
notice that $\cms(Y,X)$ is a complete metric space with the $\sup$-metric $d_{Y,X}(m,n) = \sup_{y \in Y}\{d_X(my, ny)\}$; it is then easy to show that $\Phi_f$ is an $r$-contractive map, and so its unique fixpoint is a unique non-expansive map $\sol f: Y \to X$ such that~\refeq{eq:fixp} commutes. 
      
    \item \label{ex:toptrees} Let $\cat A$ be a category with finite products, and
      let $\C$ be the presheaf category $\presh{\omega}{\A} \deq \A^{\omega^\op}$ of
      $\omega^{\mathsf{op}}$-chains in $\A$. The delay functor $\ibox$ takes a
      presheaf $X: \omega^\op \to \A$ to the presheaf $\ibox X$ with
      $\ibox X(0) = 1$ and $\ibox X(n+1) = X(n)$ for $n \geq 0$. And
      $p_X$ is given by $(p_X)_0 : X(0) \to 1$ unique and $(p_X)_{n+1}
      = X (n+1 \geq n): X(n+1) \to X(n)$. For
      every $f: \ibox X \times Y \to X$ there is a unique $\sol f: Y
      \to X$ making~\refeq{eq:fixp} commutative; it is defined
      as follows: given $f: \ibox X \times Y \to X$ (i.e.\ $f_0: Y(0)
      \to X(0)$ and $f_{n+1}: X(n) \times Y(n+1)\to X(n+1)$) one defines $\sol
      f: Y \to X$ by $\sol f_0 = f_0: Y(0) \to X(0)$ and 
      \[
      \sol f_{n+1}
      = 
      (\xymatrix@1{
        Y(n+1) \ar[rrrr]^-{\langle \sol f_n \cdot Y(n+1 \geq n),
          Y(n+1)\rangle}
        &&&&
        X(n) \times Y(n+1)
        \ar[r]^-{f_{n+1}}
        &
        X(n+1)
      }).
      \]
      It is not difficult to prove that $\sol f$ is the unique
      morphism such that~\refeq{eq:fixp} commutes. 

      Notice that for $\A = \setc$, $\C$ is the ``topos of trees'' studied by Birkedal et
      al.~\cite{BirkedalMSS12:lmcs}; they prove in Theorem~2.4 that
      $\setc^{\omega^\op}$ has a unique guarded fixpoint operator. 
      
      The next example generalizes this one. 

    \item \label{ex:presheaves} 
    Assume $\WG \deq  (W, <)$ is a well-founded poset, i.e, contains no infinite descending chains; for simplicity, we can assume $\WG$ has a root $r$. Furthermore, let $\catD$ be a (small) complete category and  $\catC \deq \presh{\WG}{\D}$, i.e.,  $\catC = \catD^{(W, >)}$. Define $(\ibox X)(w)$ to be the limit of the diagram whose nodes are $X(u)$ for $u < w$ and whose arrows are restriction morphisms: $\ibox X(w) = \lim_{v < w} X(v)$. Then as $X(w)$ itself with restriction mappings forms a cone on that diagram, a natural $p_X: X \to \ibox X$ is given by the universal property of the limits. Note that for $r$, we have that $(\ibox X)(r)$ is the terminal object $1$ of $\catD$. The $\dagger$-operation is defined as follows: given $f: \ibox X \times Y \to X$ one defines $\sol f: Y \to X$ by induction on $(W, <)$; for the root $r$ let $\sol f_r = f_r: Y(r) = 1\times Y(r) \to X(r)$, and assuming that $\sol f_v$ is already defined for all $v < w$ let
\[
\sol f_w = (\xymatrix@1{
  Y(w) \ar[rr]^-{\langle k, Y(w)\rangle} && \ibox X(w) \times Y(w) \ar[r]^-{f_w} & X(w)
}),
\]
where $k: Y(w) \to \ibox X(w)$ is the morphism uniquely induced by the cone $\sol f_v \cdot Y(w > v): Y(w) \to Y(v) \to X(v)$ for every $v < w$. One can prove that $\sol f$ is a morphism of presheaves and that it is the unique one such that~\refeq{eq:fixp} commutes. Details will be given in  the full version. Regarding the examples given in \cite{BirkedalMSS12:lmcs}, see also Proposition \ref{prop:mgrt} below.

    \item \label{ex:cpo} Let $\ibox$ be the lifting functor $(-)_\bot$ on $\cpo$,
      i.e.\ for any cpo $X$, $X_\bot$ is the cpo with a newly added
      least element. The natural transformation $p_X: X \to X_\bot$ is
      the embedding of $X$ into $X_\bot$. Then $\cpo$ has a guarded fixpoint operator given
      by taking least fixpoints. To see this notice that the hom-sets
      $\cpo(X,Y)$ are cpos with the pointwise order: $f \leq g$
      iff $f(x) \leq g(x)$ for all $x \in X$. Now any continuous $f:
      X_\bot \times Y \to X$ gives rise to a continuous
      map $\Phi_f$ on $\cpo(Y, X_\bot)$:
      \[
      \Phi_f: \cpo(Y, X_\bot) \to \cpo(Y, X_\bot), \qquad
      \Phi_f(m) = p_X \cdot f \cdot\langle m, Y\rangle.
      \]
      Using the least fixpoint $s$ of $\Phi_f$ one then defines:
      \[
      \sol f = (\xymatrix@1{
        Y 
        \ar[r]^-{\langle s, Y\rangle}
        &
        X_\bot \times Y
        \ar[r]^-f
        &
        X
      });
      \]
      using that $s = \Phi_f(s)$ it is not difficult to prove that $\sol f$ makes~\refeq{eq:fixp} commutative.
      \tlnt{GENERALIZE TO ARBITRARY FIXPOINT MONAD}\\
      \smnt{There is a problem with $\dagger\dagger$-identity.}
  \end{enumerate}
\end{examples}

Birkedal et al. \cite{BirkedalMSS12:lmcs} provide a general setting for topos-theoretic examples like \refeq{ex:toptrees} and \refeq{ex:presheaves} (the latter restricted to the case of $\setc$-presheaves) by defining a notion of \emph{a model of guarded recursive terms} and showing that \emph{sheaves over complete Heyting algebras with a well-founded basis} proposed by \cite{DiGianantonioM04:fossacs} are instances of this notion.  The difference between Definition 6.1 in  \cite{BirkedalMSS12:lmcs} and our  Definition \ref{def:dagger} is that in the former a) the delay endofunctor $\ibox$ is also assumed to preserve finite limits. On other hand b) our equality \refeq{eq:fixp} is only postulated  in the case when $Y$ is the terminal object, i.e.,  only non-parametrized fixpoint identity is assumed but c) the dagger in this less general version of \refeq{eq:fixp} is assumed to be unique.  Now, one can show that assumptions a) and c) imply our parametrized identity \refeq{eq:fixp}  \emph{whenever the underlying category is cartesian closed}, in particular whenever $\C$ is a topos.  Let us state both the definition and the result formally:

\begin{definition}[\cite{BirkedalMSS12:lmcs}]
  \label{def:mgrt} 
  A \emph{model of guarded fixpoint terms} is a triple $(\catC, \ibox, \unpdg)$, where 
  \begin{itemize}
  \item  $(\catC, \ibox)$  satisfy our general Assumption \ref{mainassumption}, i.e., $\ibox: \C \to \C$ is a pointed endofunctor (with point $p: \Id \to \ibox$) and $\catC$ has finite limits
   \item $\ibox$ preserves finite limits and
  \item  $\unpdg$ is a family of
  operations $\unpdg_{X} : \catC(\ibox X, X) \to \catC(1,X)$
  such that for every $f: \ibox X \to X$, $\ssol f$ is a unique morphism making the following square
  commute:
    \begin{equation}\label{eq:mgrt}
      \vcenter{
        \xymatrix@C+1pc{
          1 
          \ar[r]^-{\ssol f}
          \ar[d]_{\ssol f }
          &
          X
          \\
          X 
          \ar[r]_-{p_X}
          &
          \ibox X
          \ar[u]_{f}
        }
      }
    \end{equation}
\end{itemize}
\end{definition}


We write  $\iprod{X}{Y} : \ibox X \times \ibox Y \to \ibox(X \times Y)$
for the isomorphism provided by the assumption of limit preservation for the special case of product\footnote{One can note here that for the purpose of stating and proving Proposition \ref{prop:mgrt}, the assumption of finite limit preservation in Definition \ref{def:mgrt} can be weakened to finite product preservation. We only keep the stronger assumption for full consistency with \cite[Definition 6.1]{BirkedalMSS12:lmcs}.} of $X$ and $Y$. 
 
\begin{proposition} \label{prop:mgrt}
If $(\catC, \ibox, \unpdg)$ is a model of guarded recursive terms and $\catC$ is \emph{cartesian closed} with 

\begin{tabular}{>{$}l<{$}@{\;$:$\;}>{$}l<{$}>{$}l<{$}}
\curry{X}{Y}{Z}  & \catC(X \times Y, Z) & \to \catC(X, Z^Y), \\
\uncurry{X}{Y}{Z}  & \catC(X, Z^Y) & \to \catC(X \times Y, Z), \\
\eval{Y}{Z}   & Y \times Z ^Y & \to Z,
\end{tabular}

\noindent
then the operator $\dagger_{X,Y} : \catC(\ibox X \times Y, X) \to \catC(Y,X)$ defined as 
\[
\uncurry{1}{Y}{X}([\curry{\ibox(X^Y)}{Y}{X}(f \cdot \langle (\ibox \eval{Y}{X}) \cdot \iprod{Y}{X^Y} \cdot (p_Y \times \ibox(X^Y)), \prl \rangle)]^{\unpdg})
\]
is a guarded fixpoint operator on $(\catC, \ibox)$.
\end{proposition}

Obviously, we implicitly identified $Y$ and $1 \times Y$ above. Note that the converse implication does not hold. Example \ref{ex:cats}.\ref{ex:cpo} is a a guarded fixpoint category, but $(-)_\bot$ clearly fails to preserve even finite products and hence it does not yield  a model of guarded recursive terms. 

Also, while we do not have a counterexample at the moment, Proposition  \ref{prop:mgrt} is not likely to hold when the assumption that $\catC$ is cartesian closed is removed: we believe there are examples of models of guarded recursive terms which are not guarded fixpoint categories. However, to apply Proposition  \ref{prop:mgrt}, it is enough that $(\catC, \ibox, \unpdg)$ is a \emph{full subcategory} of a cartesian closed model of guarded recursive terms such that, moreover, the inclusion functor preserves products and $\ibox$.


\begin{remark} \label{rem:monads}
Monads provide perhaps the most natural and well-known examples of  pointed endofunctors.  The reader may ask whether delay endofunctors in Example \ref{ex:cats} happen to be monads.  Clearly, the delay functors in~\refeq{ex:identity}, \refeq{ex:constant} and \refeq{ex:cpo} are. In fact, while the first two ones are rather trivial monads, \ref{ex:cpo} is a paradigm example of a \emph{fixpoint monad} of Crole and Pitts \cite{cp92}.   In \refeq{ex:cms}, i.e.~the $\cms$ example, the type $\ibox\ibox A \to \ibox A$ is still inhabited (by any constant mapping), but one can easily show that monad laws cannot hold whatever candidate for monad multiplication is postulated. In the remaining (i.e., topos-theoretic) examples, monad laws fail more dramatically: $\ibox\ibox A \to \ibox A$ is not even always inhabited.  The following section discusses perhaps the most interesting subclass of monads which happen to be delay endofunctors with unique dagger. 
\end{remark}

\subsection{Completely Iterative Theories}\smnt{Added new subsection.}
\label{sec:cim}

In this subsection we will explain how categories with guarded
fixpoint operator capture a classical setting in which guarded
recursive definitions are studied---Elgot's (completely) iterative
theories~\cite{elgot75,ebt78}. The connection to guarded fixpoint
operators is most easily seen if we consider monads in lieu of Lawvere
theories, and so we follow the presentation of \emph{(completely)
  iterative monads} in~\cite{m05}. The motivating example for completely
iterative monads are infinite trees on a signature, and we 
recall this now. Let $\Sigma$ be a
signature, i.e.\ a sequence $(\Sigma_n)_{n < \omega}$ of sets of operation symbols with
prescribed arity $n$. A $\Sigma$-tree $t$ on a set $X$ of generators is a
rooted and ordered (finite or infinite) tree
whose nodes with $n>0$ children are labelled by $n$-ary operation
symbols from $\Sigma$ and a leaf is labelled by a constant symbol
from $\Sigma_0$ or by a generator from $X$. One considers systems of mutually recursive
equations of the form
\[
x_i \approx t_i(\vec x, \vec y) \qquad i \in I,
\]
where $X = \{x_i \mid i\in I\}$ is a set of recursion variables and
each $t_i$ is a $\Sigma$-tree on $X+Y$ with $Y$ a set of parameters
(i.e.\ generators that do not occur on the left-hand side of a
recursive equation). A system of recursive equations is \emph{guarded}
if none of the trees $t_i$ is only a recursion variable $x \in
X$. Every guarded system has a unique \emph{solution}, which assigns
to every recursion variable $x_i \in X$ a $\Sigma$-tree $\sol t_i(\vec
y)$ on $Y$ such that $\sol t_i(\vec y) = t_i[\sol{\vec t}(\vec y) /
\vec x]$, i.e. $t_i$ with each $x_j$ replaced by $\sol t_j(\vec
y)$. For a concrete example, let $\Sigma$ consist of a binary
operation symbol $\ast$ and a constant symbol $c$, i.e.\ $\Sigma_0 =
\{c\}$, $\Sigma_2 = \{\ast\}$ and $\Sigma_n = \emptyset$ else. Then
the following system
\[
  x_1 \approx  x_2 \ast y_1 \qquad x_2 \approx (x_1 \ast y_2) \ast c,
\]
where $y_1$ and $y_2$ are parameters, has the following unique solution:
\[
\sol t_1 = 
\vcenter{
\xy
\POS (000,000) *{\ast} = "n1"
   , (-05,-05) *{\ast} = "n2"
   , (-10,-10) *{\ast} = "n3"
   , (-15,-15) *{\ast} = "n4"
   , (-20,-20) *{\ast} = "n5"
   , (-25,-25) *{\ast} = "n6"
   , (-30,-30) *{\ast} = "n7"
   , (-35,-35) = "n8"
\POS (005,-5) *{y_1}  = "m1"
   , (000,-10) *{c} = "m2"
   , (-05,-15) *{y_2} = "m3"
   , (-10,-20) *{y_1} = "m4"
   , (-15,-25) *{c} = "m5"
   , (-20,-30) *{y_2} = "m6"
   , (-25,-35) *{y_1} = "m7"
\POS "n1" \ar @{-} "n2"
\POS "n2" \ar @{-} "n3"
\POS "n3" \ar @{-} "n4"
\POS "n4" \ar @{-} "n5"
\POS "n5" \ar @{-} "n6"
\POS "n6" \ar @{-} "n7"
\POS "n7" \ar @{.} "n8"
\ar@{-} "n1";"m1"
\ar@{-} "n2";"m2"
\ar@{-} "n3";"m3"
\ar@{-} "n4";"m4"
\ar@{-} "n5";"m5"
\ar@{-} "n6";"m6"
\ar@{-} "n7";"m7"
\endxy
}
\qquad \textrm{and}\qquad 
\sol t_2 = 
\vcenter{
\xy
\POS (000,000) *{\ast} = "n1"
   , (-05,-05) *{\ast} = "n2"
   , (-10,-10) *{\ast} = "n3"
   , (-15,-15) *{\ast} = "n4"
   , (-20,-20) *{\ast} = "n5"
   , (-25,-25) *{\ast} = "n6"
   , (-30,-30) *{\ast} = "n7"
   , (-35,-35) = "n8"
\POS (005,-5) *{c}  = "m1"
   , (000,-10) *{y_2} = "m2"
   , (-05,-15) *{y_1} = "m3"
   , (-10,-20) *{c} = "m4"
   , (-15,-25) *{y_2} = "m5"
   , (-20,-30) *{y_1} = "m6"
   , (-25,-35) *{c} = "m7"
\POS "n1" \ar @{-} "n2"
\POS "n2" \ar @{-} "n3"
\POS "n3" \ar @{-} "n4"
\POS "n4" \ar @{-} "n5"
\POS "n5" \ar @{-} "n6"
\POS "n6" \ar @{-} "n7"
\POS "n7" \ar @{.} "n8"
\ar@{-} "n1";"m1"
\ar@{-} "n2";"m2"
\ar@{-} "n3";"m3"
\ar@{-} "n4";"m4"
\ar@{-} "n5";"m5"
\ar@{-} "n6";"m6"
\ar@{-} "n7";"m7"
\endxy}
\]

For any set $X$, let $T_\Sigma(X)$ be the set of $\Sigma$-trees on
$X$. It has been realized by Badouel~\cite{badouel89} that $T_\Sigma$
is the object part of a monad. A system of equations is then nothing
but a map
\[
f: X \to T_\Sigma(X+Y)
\]
and a solution is a map $\sol f: X \to T_\Sigma Y$ such that the
following square commutes:
\[
\xymatrix@C+2pc{
  X 
  \ar[r]^-{\sol f}
  \ar[d]_f
  &
  T_\Sigma Y
  \\
  T_\Sigma(X+Y) 
  \ar[r]_-{[\sol f, \eta_Y]}
  &
  T_\Sigma T_\Sigma Y
  \ar[u]_{\mu_Y}
  }
\]
where $\eta$ and $\mu$ are the unit and multiplication of the monad
$T_\Sigma$, respectively. 

It is clear that the notion of equation and solution can be formulated
for every monad $S$. However, the notion of guardedness requires one
to speak about \emph{non-variables} in $S$. This is enabled by Elgot's
notion of \emph{ideal theory}~\cite{elgot75}, which for a finitary monad on $\setc$ is
equivalent to the notion recalled in the following definition. We
assume for the rest of this subsection that $\cat A$ is a category
with finite coproducts such that coproduct injections are
monomorphic. 

\begin{definition}[\cite{aamv03}]
  By an {\em ideal monad} on $\A$ is understood a six-tuple
  \begin{displaymath}
    (S,\eta,\mu,S',\sigma,\mu')
  \end{displaymath}
  consisting of a monad $(S,\eta,\mu)$ on $\A$, a subfunctor $\sigma:S'\subto S$
  and a natural transformation $\mu':S'S\to S'$ such that 
  \begin{enumerate}[(1)]
  \item $S=S'+\Id$ with coproduct injections $\sigma$ and $\eta$, and
  \item $\mu$ restricts to $\mu'$ along $\sigma$, i.e., 
    the square below commutes:
    \begin{displaymath}
      \xymatrix{
        S'S
        \ar[0,2]^-{\mu'}
        \ar[1,0]_{\sigma S}
        &
        &
        S'
        \ar[1,0]^{\sigma}
        \\
        SS
        \ar[0,2]_-{\mu}
        &
        &
        S
      }
    \end{displaymath} 
  \end{enumerate}
\end{definition}
The subfunctor $S'$ of an ideal monad $S$ allows us to formulate the
notion of a guarded equation system abstractly; this leads to the
notion of completely iterative theory of Elgot et al.~\cite{ebt78} for
which we here present the formulation with monads from~\cite{m05}:
\begin{definition}
  \label{def:eqsol}
  Let $(S,\eta,\mu,S',\sigma, \mu')$ be an ideal monad on $\A$.
  \begin{enumerate}
  \item By an {\em equation morphism} is meant a morphism
    \begin{displaymath}
      f:X\to S(X+Y)
    \end{displaymath}
    in $\A$, where $X$ is an object
    (``of variables'') and $Y$ is an object (``of parameters'').
\item By a {\em solution} of $f$ is meant a morphism $\sol{f}:X\to SY$
  for which the following square commutes:
  \begin{equation}\label{diag:solmon}
    \vcenter{
    \xymatrix@C+2pc{
      X
      \ar[r]^-{\sol{f}}
      \ar[d]_{f}
      &
      SY
      \\
      S(X+Y)
      \ar[r]_-{S[\sol{f},\eta_Y]}
      &
      SSY
      \ar[u]_{\mu_Y}
      }}
  \end{equation}
\item The equation morphism $f$ is called {\em guarded} if it factors
  through the summand $S'(X+Y) + Y$ of $S(X+Y) = S'(X+Y) + X + Y$:
  \begin{displaymath}
    \xymatrix@C+2pc{
      X
      \ar[r]^-{f}
      \ar @{.>} [rd]
      &
      S(X+Y)
      \\
      &
      S'(X+Y)+Y
      \ar[-1,0]_{[\sigma_{X+Y},\eta_{X+Y}\cdot\inr]}
      }
  \end{displaymath}      
\item The given ideal monad is called \emph{completely iterative} if every guarded
  equation morphism has a unique solution. 
\end{enumerate}
\end{definition}

\begin{examples}
  We only briefly mention two examples of completely iterative
  monads. More can be found in~\cite{aamv03,m05,am06}. 
  \begin{enumerate}[(1)]
  \item
    The monad $T_\Sigma$ of $\Sigma$-trees is a completely iterative
    monad.  
  \item A more general example is given by parametrized final
    coalgebras. Let $H: \A \to \A$ be an endofunctor such that for
    every object $X$ of $\A$ a final coalgebra $TX$ for $H(-) + X$
    exists. 
    Then $T$ is the object assignment of a completely iterative
    monad; in fact, $T$ is the free completely iterative monad on
    $H$ (see~\cite{m05}). 
  \end{enumerate}
\end{examples}

We will now explain how completely iterative monads are subsumed by
the notion of categories with a guarded fixpoint operator. To this end
we fix a completely iterative monad $S$. We will show that the dual of
its Kleisli category $\C = (\A_S)^{op}$ is equipped with a guarded
fixpoint operator. First notice, that since $\A_S$ has coproducts
given by the coproducts in $\A$ we see that $\C$ has products. Next we
need to obtain the endofunctor $\ibox$ on $\C$. This will be given as
the dual of an extension of the subfunctor $S': \A \to \A$ of $S$ to
the Kleisli category $\A_S$. Indeed, it is well-known that to have an
extension of $S'$ to $\A_S$ is equivalent to having a distributive law
of the functor $S'$ over the monad $S$ (see Mulry~\cite{mulry94}).

But it is easy to verify that the natural transformation
\[
\xymatrix@1{S'S \ar[r]^-{\mu'} & S' \ar[r]^-{\eta S'} & SS'}
\]
satisfies the two required laws and thus yields a distributive
law. Moreover, the ensuing endofunctor $\ibox^{op} = S'$ on $\A_S$ is
copointed, i.e.\ we have a natural transformation $p$ from $S'$ to $\Id: \A_S
\to \A_S$; indeed, its components at $X$ are given by the coproduct injections
$\sigma_X: S'X \to SX$, and it is not difficult to verify that this
is a natural transformation; thus, $\ibox$ is a pointed endofunctor on
$\C$. 

Now observe that a morphism $f: \ibox X \times Y \to X$ is
equivalently a morphism
\[
f: X \to S(S'X + Y)
\]
in $\A$. We are ready to describe the guarded fixpoint operator on $\C$. 

\begin{construction}
  \label{con:dagger}
  For any morphism $f: X \to S(S'X + Y)$ form the following morphism
  \[
  \ol f = (\xymatrix@1{
    X \ar[r]^-{f} & S(S'X + Y) \ar[rr]^-{S(\sigma_X + \eta_Y)} &&
    S(SX + SY) \ar[r]^-{S\can} & SS(X+Y) \ar[r]^-{\mu_{X+Y}} & S(X+Y)
  }),
  \]
  where $\can = [S\inl,S\inr]: SX + SY \to S(X+Y)$. 
  It is not difficult to verify that $\ol f$ is a guarded equation
  morphism for $S$, and we define $\sol f: X \to SY$ to be the unique
  solution of $\ol f$. 
\end{construction}

\begin{proposition}
  For every $f$, $\sol f$ from Construction~\ref{con:dagger} is a
  unique morphism $Y \to X$ in $\C$ such that~\refeq{eq:fixp}
  commutes. 
\end{proposition}

In fact, to prove this proposition one shows that solutions of $\ol f: X \to
S(X+Y)$ (i.e.\ morphisms $s: X \to SY$ such that~\refeq{diag:solmon}
commutes) are in one-to-one correspondence with morphisms $Y\to X$ is
$\C$ such that~\refeq{eq:fixp} commutes.

\subsection{Properties of Guarded Fixpoint Operators}
\label{sec:prop}

In this section we study properties of
guarded fixpoint operators. Except for uniformity these properties are
purely equational. They are generalizing analogous 
properties of iteration theories; more precisely, they would collapse to the original, unguarded counterparts when $\ibox$ is instantiated to the identity endofunctor (see Example~\ref{ex:cats}(\ref{ex:identity})). 

\begin{definition}
  \label{def:prop}
  Let $(\catC,\ibox, \dagger)$ be a guarded fixpoint category. We define the following properties of $\dagger$:

  \begin{enumerate}[{\bf (1)}]
  \item {\bf Fixpoint Identity.} For every $f: \ibox X \times Y \to X$ the
    diagram~\refeq{eq:fixp} commutes. This is built into the definition of guarded fixpoint categories and only mentioned here again for the sake of completeness.
  \item {\bf Parameter Identity.} For every $f: \ibox X \times Y \to X$ and
    every $h: Z \to Y$ we have
    \[
    \xymatrix@1{
      Z \ar[r]^-h & Y \ar[r]^{\sol f} & X
    }
    =
    (\xymatrix@1@C+1pc{
      \ibox X \times Z
      \ar[r]^-{\ibox X \times h}
      &
      \ibox X \times Y
      \ar[r]^-f
      &
      X
    })^\dagger.
    \]
    
      \ifbool{full}{\[\inferrule{\Gamma, x:\ibox X \vdash F: X  \quad \sigma \text{ a substitution of typed variables leaving } x \text{ unchanged}}{\Gamma \vdash \sigma(\fix{x}{F}) = \Gamma \vdash \fix{x}{\sigma F}}
    \]
    \tlnt{I'm not sure---this is a tentative version}}
  \item {\bf (Simplified) Composition Identity.} Given $f:\ibox X \times Y
    \to Z$ and $g: Z \to X$ we have
    \[
    (\xymatrix@1{
      \ibox X \times Y \ar[r]^-{f} & Z \ar[r]^-g & X
    })^\dagger
    =
    (\xymatrix@1{
      Y \ar[rr]^-{(f \cdot (\ibox g \times Y))^\dagger} && Z \ar[r]^-{g} & X
      }).
    \]
    \ifbool{full}{\[
    \inferrule{\Gamma, x: \ibox X \vdash F: Z \qquad z: Z \vdash G: X}{\Gamma \vdash \fix{x}{G[F/z]} = G[F[\comono{G}{z}{z'}/x]/z]}
    \]}{}
 
  \item {\bf Double Dagger Identity.} For every $f: \ibox X \times \ibox X
    \times Y \to X$ we have
    \[
    (\xymatrix@1{
      Y \ar[r]^{f^{\dagger\dagger}} & X
    }) 
    =
    (\xymatrix@1{
      \ibox X \times Y \ar[r]^-{\Delta \times Y}
      &
      \ibox X \times \ibox X \times Y \ar[r]^-f & X
    })^\dagger.
    \]
 \ifbool{full}{\[
 \inferrule{\Gamma, x_1:\ibox X,x_2:\ibox X \vdash F: X}{\Gamma \vdash \fix{x_1}{\fix{x_2}{F}} = \fix{x}{F[x_2/x_1]}}
 \]  } 
  \item {\bf Uniformity.} Given $f: \ibox X \times Y \to X$, $g: \ibox X'
    \times Y \to X'$ and $h: X\to X'$ we have
    \[
    \vcenter{
      \xymatrix{
        \ibox X \times Y \ar[r]^-f \ar[d]_{\ibox h \times Y} & X \ar[d]^h \\
        \ibox X' \times Y \ar[r]_-g & X'
      }}
    \qquad
    \implies
    \qquad
    \vcenter{
      \xymatrix@R-1pc{
        & X \ar[dd]^h \\
        Y
        \ar[ru]^-{\sol f}
        \ar[rd]_-{\sol g}
        \\
        & X'
      }}
    \]
  \end{enumerate}
\ifbool{full}{  
\[
\inferrule{\Gamma, x: \ibox X \vdash F: X \quad \Gamma, x' : X' \vdash G: X, x: X \vdash H: X' \quad \Gamma \vdash G[\comono{H}{x}{x''}/x] = H[F/x] }{\fix{x'}{G} = H[\fix{x}{F}/x]}
\]}{}

  We call the first four properties (1)--(4) the \emph{Conway}
  axioms. 
\end{definition}

Notice that the Conway axioms are equational properties while (5) is
quasiequational (i.e.\ an implication between equations).

Next we shall show that in the presence of certain of the above
properties the natural transformation $p: \Id \to \ibox$ is a derived
structure. Let $(\catC,\ibox)$ be equipped with an operator $\dagger$ 
\emph{not necessarily
satisfying~\refeq{eq:fixp}}. For every object $X$ of $\catC$ 
define $q_X: X \to \ibox X$ as follows: consider
\[
f_X = (\xymatrix@1@C+1.5pc{
  \ibox(\ibox X \times X) \times X \ar[r]^-{\ibox \prr \times X} &
  \ibox X \times X
})
\]
and form
\[
q_X = (\xymatrix@1{
  X \ar[r]^-{f_X^\dagger} & \ibox X \times X \ar[r]^-{\prl} & \ibox X
}).
\]

\begin{lemma}
  Let $(\catC,\ibox)$ be equipped with the operator $\dagger$. Then:
  \begin{enumerate}
  \item If $\dagger$ satisfies the parameter identity and uniformity,
    then $q: \Id \to \ibox$ is a natural transformation.
  \item If $\dagger$ satisfies the fixpoint identity, then $q_X =
    p_X$ for all $X$. 
  \end{enumerate}
\end{lemma}

\begin{definition}
   A guarded fixpoint category $(\catC,\ibox, \dagger)$  satisfying
  the Conway axioms (i.e.\ fixpoint, parameter, composition and double dagger identities) is called a \emph{guarded Conway category}.

  If in addition 
  uniformity is satisfied, we call $(\catC,\ibox,\dagger)$ a \emph{uniform guarded Conway
  category}. 

  And $(\catC,\ibox,\dagger)$ is called a \emph{unique guarded fixpoint category} if
  for every $f: \ibox X \times Y \to X$, 
  $\sol f: Y \to X$ is the unique morphism such that~\refeq{eq:fixp} commutes. In this case, we can just write a pair $(\catC,\ibox)$ rather than a triple $(\catC,\ibox,\dagger)$.
\end{definition}

The next theorem states that such a unique $\dagger$ satisfies all
the properties in Definition~\ref{def:prop}. 

\begin{theorem}
  \label{thm:unique}
  If $(\catC,\ibox)$ is a unique guarded fixpoint category, then it is a
  uniform guarded Conway category. 
\end{theorem}
%
%
\begin{examples}
  \begin{enumerate}[(1)]
  \item Several of our examples in~\ref{ex:cats} are unique guarded fixpoint
    categories and hence their unique $\dagger$ satisfies all the
    properties in Definition~\ref{def:prop}. This holds for
    Examples~\ref{ex:cats}(2)--(6), and also for the example of completely iterative monads in Section~\ref{sec:cim}. 
  \item One can prove that Example~\ref{ex:cats}(7), i.e.,\ $\catC = \cpo$
    with the lifting functor $\ibox = (-)_\bot$ satisfies all the
    properties of Definition~\ref{def:prop}, i.e.\ $(\cpo, (-)_\bot)$ is a uniform guarded Conway
    category. But it is not a unique guarded fixpoint category: for
    let $X = \{0,1\}$ be the two-chain, $Y = 1$ the one element cpo
    and $f: X_\bot = X_\bot \times Y \to X$ be the map with $f(0) =
    f(\bot) = 0$ and $f(1) = 1$. Then both $0: 1 \to X$ and $1: 1 \to
    X$ make~\refeq{eq:fixp} commutative.
  \end{enumerate}
\end{examples}

\ifbool{full}{
\subsection{Fixpoint expressions on types}

The starting point the fixpoint theorem in \cite{Sambin76:sl,Visser05:lncs} \dots For  importance of such results in contemporary type theory, see \cite{Nakano00:lics,BirkedalMSS12:lmcs,AtkeyMB13:icfp} \dots}{}

\section{Guarded Trace Operators}
\label{sec:tr}

In the case special case where $\ibox$ is the identity functor (see
Example~\ref{ex:cats}(1)), it is well-known that a fixpoint operator 
satisfying the Conway axioms is equivalent to a trace operator w.r.t.\ the product on $\catC$ (see
Hasegawa~\cite{h97,h99}). In this section we present a similar
result for a 
generalized notion of a guarded trace operator on $(\catC, \ibox)$. 

\begin{remark}
Recall that the notion of an (ordinary) trace operator was introduced by
Joyal, Street and Verity~\cite{jsv96} for symmetric monoidal
categories. The applicability of the notion of trace to non-cartesian tensor products is in fact one of main reasons of its popularity.  Our generalization can also be
formulated for symmetric monoidal categories, see the remark preceding Construction \ref{con:inv} below. However, the main results in this section, i.e., Theorems \ref{thm:tr} and \ref{thm:unif} do not make any use of this added generality. Hence, we keep the Assumption \ref{mainassumption} like in the remainder of the paper.
\end{remark}

\begin{definition}
  A (cartesian) \emph{guarded trace operator} on $(\catC, \ibox)$ is a natural
  family of operations
  \[
  \Tr_{A,B}^X: \catC(\ibox X \times A, X \times B) \to \catC(A,B)
  \]
  subject to the following three conditions:
  \begin{enumerate}
  \item {\bf Vanishing.} (I) For every $f: \ibox 1 \times A \to B$ we
    have
    \[
    \Tr_{A,B}^1(f) = (\xymatrix@1@C+1pc{
      A \cong 1 \times A
      \ar[r]^-{p_1 \times A}
      &
      \ibox 1 \times A
      \ar[r]^-f
      &
      B
      }).
    \]
    (II) For every $f: \ibox X \times \ibox Y \times A \to X\times Y
    \times B$ we have
    \[
    \small
    \Tr_{A,B}^Y(\Tr_{\ibox Y \times A, Y \times A}^X(f)) =
    \Tr_{A,B}^{X\times Y} 
    (\xymatrix@1@C-.5pc{
      \ibox(X \times Y) \times A \ar[rr]^-{\can \times A}
      &&
      \ibox X \times \ibox Y \times A \ar[r]^-f
      &
      X \times Y \times A
      }).
    \]
  \item {\bf Superposing.} For every $f: \ibox X \times A \to X \times B$ we
    have
    \[
    \Tr_{A \times C, B \times C}^X (f \times C) = \Tr_{A,B}^X(f)
    \times C.
    \]
  \item {\bf Yanking.} Consider the canonical isomorphism $c: \ibox X \times
    X \to X \times \ibox X$. Then we have
    \[
    \Tr_{X,\ibox X}^X(c) = (\xymatrix@1{X\ar[r]^{p_X} & \ibox X}).
    \]
  \end{enumerate}
  
  If $\Tr$ is a (cartesian) guarded trace operator on $(\catC, \ibox)$, $(\catC, \ibox, \Tr)$ is called
  a \emph{guarded traced (cartesian) category}.
\end{definition}

Of course, when $\ibox$ is taken to be the identity on $\C$ (as in Example~\ref{ex:cats}(\ref{ex:identity})), our notion of guarded trace specializes to the notion of an ordinary trace operator (w.r.t.~product) of Joyal, Street and Verity. 

In addition, as in the case of ordinary trace operators naturality of $\Tr$ can
equivalently be expressed by three more axioms:
\begin{enumerate}
  \setcounter{enumi}{3}
\item {\bf Left-tightening.} Given $f: \ibox X \times A \to X \times
  B$ and $g: A' \to A$ we have
  \[
  \Tr_{A',B}^X(\xymatrix@1{
    \ibox X \times A'
    \ar[rr]^-{\ibox X \times g}
    &&
    \ibox X \times A 
    \ar[r]^-f
    &
    X \times B
  })
  =
  (\xymatrix@1{
    A'\ar[r]^-g 
    & 
    A \ar[rr]^-{\Tr_{A,B}^X(f)}
    &&
    B
  }).
  \]
\item {\bf Right-tightening.} Given $f: \ibox X \times A \to X \times
  B$ and $g: B \to B'$ we have
  \[
  \Tr_{A,B'}^X(\xymatrix@1{
    \ibox X \times A \ar[r]^-f 
    &
    X \times B
    \ar[r]^-{X \times g}
    &
    X \times B'
  }) 
  = 
  (\xymatrix@1{
    A \ar[rr]^-{\Tr_{A,B}^X(f)} && B \ar[r]^-g & B'
  }).
  \]
\item {\bf Sliding.} Given $f: \ibox X \times A \to X' \times B$ and
  $g: X' \to X$ we have
  \[
  \Tr_{A,B}^X(\xymatrix@1{
    \ibox X \times A \ar[r]^-f 
    & 
    X' \times B
    \ar[r]^-{g \times B}
    &
    X \times B
  })
  =
  \Tr_{A,B}^{X'}(\xymatrix@1{
    \ibox X' \times A \ar[r]^{\ibox g \times A}
    &
    \ibox X \times A
    \ar[r]^-f
    &
    X ' \times B
  }).
  \]
\end{enumerate}

\begin{remark}
  The generalization for a symmetric monoidal category $(\catC, \otimes, I,
  c)$ equipped with a pointed endofunctor $\ibox: \cat C \to \catC$
   requires the assumption that $\ibox$ is \emph{comonoidal}, i.e., equipped with a morphism
  $m_I: \ibox I \to  I$ and a natural transformation $m_{X,Y}: \ibox (X
  \times Y) \to \ibox X \times \ibox Y$ satisfying the usual coherence
  conditions. In fact, in the formulation of Vanishing (II) we used
  that in every category the product $\times$ is comonoidal via $m_{X,Y} =
  \can$. 
\end{remark}

\iffull
\begin{example}
\tlnt{well \dots do we have any additional ones?} Suitably modified tracing on relations? Abramsky et al. on nuclear traces?
\end{example}
\fi

\begin{construction}\label{con:inv}
  \begin{enumerate}
  \item Let $(\catC, \ibox, \Tr)$ be a guarded traced category. Define a guarded
    fixpoint operator $\dace: \catC(\ibox X \times A) \to
    \catC(A,X)$ by
    \[
    f^{\dace} = \Tr_{A,X}^X(\xymatrix@1{
      \ibox X \times A \ar[r]^-{\langle f, f\rangle} & X \times X)
    }): A \to X.
    \]
  \item Conversely, suppose $(\catC,\ibox, \dagger)$ is  a guarded
    fixpoint category. Define $\tragger_{A,B}^X:
    \catC(\ibox X \times A, X \times B) \to \catC(A,B)$ by setting for
    every $f: \ibox X \times A \to X \times B$
    \[
    \tragger_{A,B}^X (f) =
    (\xymatrix@1{
      A \ar[rr]^-{\langle \sol{(\prl \cdot f)}, A\rangle}
      &&
      X \times A
      \ar[rr]^-{p_X \times A}
      &&
      \ibox X \times A
      \ar[r]^-f
      &
      X \times B
      \ar[r]^-{\prr}
      &
      B 
    }). 
    \]
  \end{enumerate}
\end{construction}

\enlargethispage{12pt}
The main result in this section states that the category $\catC$ is
guarded traced iff it is a guarded Conway category:
\begin{theorem}\label{thm:tr}
\begin{enumerate}
  \item Whenever $(\catC,\ibox, \Tr)$ is a guarded traced category,  $(\catC,\ibox,\dace)$ is a guarded Conway
  category. Furthermore, $\Tr_{\dace}$ is the original operator $\Tr$.
  \item Whenever $(\catC,\ibox, \dagger)$ is a guarded Conway category,  $(\catC,\ibox,\tragger)$ is guarded traced. Furthermore, $\dagger_{\tragger}$ is the original operator $\dagger$.
  \end{enumerate}
\end{theorem}


The proof details are similar to the proof details for ordinary
fixpoint operators and traced cartesian categories (see
Hasegawa~\cite{h99}). Here one has to stick $\ibox$ in ``all the
right places'' in all the necessary verifications of the axioms
for trace and dagger, respectively. However, some of proof steps, in particular the derivation of a guarded version  of the so-called Beki\v{c} identity require some creativity; it is not a completely automatic adaptation.

Hasegawa  related uniformity of trace to uniformity of dagger and we
can do the same in the guarded setup. Recall that in
iteration theories uniformity (called \emph{functorial dagger implication})
plays an important role. On the one hand, this quasiequation implies
the so-called \emph{commutative identities}, an infinite set of equational
axioms that are added to the Conway axioms in order to yield a complete
axiomatization of fixpoint operators in domains. On the other hand,
most examples of iteration theories actually satisfy uniformity, and
so uniformity gives a convenient sufficient condition to verify that a
given Conway theory is actually an iteration theory. 


\begin{definition}
  A guarded trace operator $\Tr$ is called uniform if for every
  morphism $f: \ibox X \times A \to X \times B$, $f': \ibox X' \times
  A \to X' \times B$ and $h: X \to X'$ we have
  \[
  \vcenter{
    \xymatrix{
      \ibox X \times A \ar[r]^-f
      \ar[d]_{\ibox h \times A}
      &
      X \times B
      \ar[d]^{h \times B}
      \\
      \ibox X' \times A \ar[r]_-{f'}
      &
      X' \times B
    }}
  \qquad\implies\qquad
  \Tr_{A,B}^X(f) = \Tr_{A,B}^{X'}(f'): A \to B.
  \]
\end{definition}

\begin{theorem}\label{thm:unif}
\begin{enumerate}
  \item Whenever $(\catC,\ibox, \Tr)$ is a uniform guarded traced category,  $\dace$ is a uniform guarded Conway
  operator.
  \item Whenever $(\catC,\ibox, \dagger)$ is a uniform guarded Conway category,  $\tragger$ is a uniform guarded trace operator. 
  \end{enumerate}
\end{theorem}


\begin{remark}
  Actually, Hasegawa proved a slightly stronger statement concerning
  uniformity then what we stated in Theorem~\ref{thm:unif}; he showed
  that a Conway operator is uniform w.r.t.~any fixed morphism $h: X
  \to X'$ (i.e.\ satisfies uniformity just for $h$) iff the
  corresponding trace operator is uniform w.r.t. this morphism $h$.  The proof is somewhat more complicated and in our guarded setting 
   we leave this as an exercise to the
  reader. 
\end{remark}

Finally, let us note that the bijective correspondence between guarded
Conway operators and guarded trace operators established in
Theorem~\ref{thm:tr} yields an isomorphism of the \mbox{(2-)}categories of
(small) guarded Conway categories and guarded traced (cartesian)
categories. The corresponding notions of morphisms are, of course, as
expected:

\begin{definition}
\
  \begin{enumerate}
  \item $F: (\C,\ibox^\C,\dagger) \to (\D,\ibox^\D,\ddagger)$ is a morphism of guarded Conway categories whenever $F:
    \C \to \D$ is a finite-product-preserving functor satisfying
    \begin{equation}\label{eq:sat}
      \vcenter{
        \xymatrix{
          \C 
          \ar[r]^{\ibox^\C}
          \ar[d]_F
          &
          \C
          \ar[d]^F
          \\
          \D \ar[r]_-{\ibox^\D}
          &
          \D
        }}
      \quad
      \text{and}
      \quad
      p^\D_{FX} = F(p_X^\C): FX \to \ibox^\D FX = F(\ibox^C X),
    \end{equation}
    and  preserving dagger, i.e., for every $f: \ibox X \times A \to X$ we
    have
    \[
    F(\sol f) = (\xymatrix@1{
      \ibox^\D FX \times FA \cong F(\ibox^\C X \times A) \ar[r]^-{Ff} & FX
    })^\ddagger.
    \]
  \item A morphism $F: (\C,\ibox^\C,\Tr_{\C}) \to (\D,\ibox^{\D},\Tr_{\D})$ 
    is a finite-product-preserving $F: \C \to \D$ 
    satisfying~\refeq{eq:sat} above and preserving the trace
    operation: for every $f: \ibox^\C X \times A \to X \times B$ in
    $\C$ we have
    \[
    F(\Tr_{\C\,A,B}^{\;\;\; X}(f)) = \Tr_{\D\,FA,FB}^{\quad\! FX}(\xymatrix@1{
      \ibox^\D FX \times FA \cong F(\ibox^\C X \times A) \ar[r]^-{Ff}
      & F(X \times B) \cong FX \times FB
    }).
    \] 
  \end{enumerate}
\end{definition}

\begin{corollary}
  The (2-)categories of guarded Conway categories and of guarded
  traced (cartesian) categories are isomorphic.
\end{corollary}

\section{Conclusions and Future Work}
\label{sec:conc}

We have made the first steps in the study of equational
properties of guarded fixpoint operators popular in the recent literature, e.g., \cite{Nakano00:lics,Nakano01:tacs,AppelMRV07:popl,BentonT09:tldi,BirkedalMSS12:lmcs,KrishnaswamiB11:lics,KrishnaswamiB11:icfp,BirkedalMSS12:lmcs,AtkeyMB13:icfp}. We began with an extensive list of examples, including both those already discussed in the above references  and some whose connection with the ``later'' modality has not seemed obvious so far---e.g., Example \ref{ex:cats}.\ref{ex:cpo} or completely iterative theories in Section \ref{sec:cim}. Furthermore, we
formulated the four Conway properties and uniformity in analogy to the
respective properties in iteration theories and we showed them to be
sound w.r.t.~all models discussed in Section \ref{sec:fix}. In
particular, Theorem \ref{thm:unique} proved that our axioms hold in all categories with
a unique guarded dagger. In Theorem \ref{thm:tr}, we have a generalization
of a result by Hasegawa for ordinary fixpoint operators: we proved
that to give a (uniform) guarded fixpoint operator satisfying the
Conway axioms is equivalent to giving a (uniform) guarded trace
operator on the same category.

Our paper can be considered as a work in progress report. 
Our aim is
to eventually arrive at completeness results similar to the ones on
iteration theories. We do not claim that the axioms
we presented are complete. In the unguarded setting, completeness is obtained by adding to the Conway axioms an
infinite set of equational axioms called the \emph{commutative
identities}, see~\cite{be93,sp00}. We did not consider those here,
but we considered the quasi-equational property of uniformity which
implies the commutative identities and is satisfied in most models of
interest. Only further research can show whether this property can ensure completeness in the guarded setup or one  needs to postulate stronger ones.

Other future work pertains to a syntactic type-theoretic presentation
of the axioms we studied and a description of a
classifying guarded Conway category. 

Concerning further models of guarded fixpoint operators, it would
be worthwhile to consider fixpoint monads of Crole and
Pitts~\cite{cp92} more closely. These generalize our example of the
category $\cpo$ with the lifting monad. One can
prove that any fixpoint monad induces a
guarded fixpoint operator satisfying parameter and simplified 
composition identities as well as uniformity. However, proving the
double dagger identity in the general case is an open problem. 

It would also be interesting to obtain examples of guarded traced monoidal categories which are not ordinary traced monoidal categories and which do not arise from guarded Conway categories. Traces w.r.t.~a trace ideal as considered by Abramsky, Blute and Panangaden~\cite{abp98} might be a good starting point.

\subparagraph*{Acknowledgements}

We would like to acknowledge an inspiring
discussion with Erwin R.~Catesbeiana  on (un-)productive (non-)termination. We would also like to thank in general William and Arthur for their very insistence on  major modal undertones in modern modelling of this phenomenon. 

\appendix

\tlnt{appendix if needed \dots}



\bibliography{intmod,continuations,guarded-dagger}


\end{document}